\let\singlecol\undefined 

\ifdefined\singlecol
  \documentclass[useAMS,usenatbib,referee]{aa}
\else
  \documentclass[useAMS,usenatbib]{aa}
\fi
\usepackage{graphicx}
\usepackage[utf8]{inputenc}
\usepackage{txfonts}
\usepackage{times}
\usepackage{longtable}
\usepackage{natbib}
\usepackage{wasysym}
\usepackage{pdflscape}
\usepackage{caption}
\usepackage{xcolor}
\usepackage[switch]{lineno}
\usepackage{hyperref}
\hypersetup{
    colorlinks=true,
    linkcolor=magenta,
    citecolor=blue,
    filecolor=magenta,      
    urlcolor=cyan,
}
\usepackage[separate-uncertainty=true]{siunitx}
\usepackage{makecell}
\usepackage{threeparttable}
\usepackage{dblfloatfix}    
\usepackage{breqn}
\usepackage{tabularx}

\defcitealias{LHAASO_PeV}{C21}
\defcitealias{Breuhaus}{B21}

\DeclareSIUnit\gauss{G}
\DeclareSIUnit\erg{erg}
\DeclareSIUnit\pc{pc}
\DeclareSIUnit\year{yrs}
\DeclareSIUnit\jansky{Jy}
\DeclareSIUnit\beam{beam}

\def\@fnsymbol#1{\ensuremath{\ifcase#1\or *\or \dagger\or \ddagger\or
   \mathsection\or \mathparagraph\or \|\or **\or \dagger\dagger
   \or \ddagger\ddagger \else\@ctrerr\fi}}

\title{A pulsar wind nebula origin of the LHAASO-detected UHE $\gamma$-ray sources}
\author{M. Breuhaus \inst{\star}, B. Reville, J. A. Hinton}

\authorrunning{Breuhaus et al.}
\institute{Max-Planck-Institut f\"ur Kernphysik, Postfach 103980, D 69029 Heidelberg, Germany}

\date{Accepted 2021-12-23; Received 2021-08-26}

\abstract{
The recent measurement by LHAASO of gamma-ray emission extending up to 100s of TeV from multiple Galactic sources represents a major observational step forward in the search for the origin of the Galactic cosmic rays. The burning question is if this ultra-high-energy emission is associated to acceleration of protons and/or nuclei to PeV energies, or can be associated to PeV-electron accelerators. A strong Klein-Nishina suppression of inverse Compton emission at these energies is unavoidable, nevertheless we show here that inverse Compton emission can provide a natural explanation of the measured emission, and that 
an association to the established PeV-electron accelerating source class of pulsar wind nebulae is also rather natural. However, a clear distinction between different models requires taking into account multi-wavelength data, having good knowledge about the local environmental conditions and, in some cases, to perform multi-source modeling.
}

\keywords{}

\begin{document}
\maketitle

\protect\footnotetext[1]{corresponding author:  Mischa Breuhaus, \email{mischa.breuhaus@mpi-hd.mpg.de}\\Member of the International Max-Planck Research School for Astronomy and Cosmic Physics at the University of Heidelberg (IMPRS-HD), Germany}

\section{Introduction}

The HAWC, Tibet AS$_\gamma$ and LHAASO collaborations have reported detection of multiple Galactic ultra-high energy  gamma-ray (UHE: $E_\gamma\geq 100$ TeV) sources \citep{HAWC_UHE, Tibet_AS_gamma_Crab_2019, Tibet_AS_gamma_Cygnus_2021, LHAASO_PeV}. The latest published data from the partially completed LHAASO observatory \citep{LHAASO_PeV} reveal 12 sources with significant ($> 7 \sigma$) detections above several 100 TeV. The completed LHAASO facility will greatly enhance our capability to observe the Galaxy's most extreme particle accelerators. 
The unprecedented sensitivity in the UHE regime provided by LHAASO and other planned facilities such as CTA \citep{CTA_sciencecase_2019} and SWGO \citep{SWGO_sciencecase}, is particularly important for efforts to identify Galactic sources of $>$ PeV cosmic rays (CRs). Acceleration to the knee energy in the CR spectrum is a long-standing issue for the supernova-remnant (SNR) CR origin theory \citep{Lagage_Cesarsky_1983,Bell2013}, and though UHE $\gamma$-rays provide the best method to identify PeV CR accelerators in our Galaxy, a critical examination of all such sources is warranted. 

Hard $\gamma$-ray spectra at these energies are generally perceived to favour hadronic emission mechanisms, since the Klein-Nishina suppression of inverse Compton (IC) emission at UHEs can, in many circumstances, disfavour hard equilibrium spectra. However, this may not apply if the emission region is moderately photon dominated \cite[e.g.][]{BG70,Aharonian_1985,Zdziarski93}.
Recently, \citet{Breuhaus} (henceforth \lq\lq\citetalias{Breuhaus}\rq\rq) have pointed out that the emission from high-power pulsars may naturally account for hard spectra UHE sources \cite[see also][]{Di_Mauro_et_al_2020,Sudoh_et_al_2021}. Observations reported by the HAWC collaboration already indicate that UHE $\gamma$-ray emission in the vicinity of pulsars with high spin-down powers of $\dot E > \SI{e36}{\erg\per\second}$ may be common \citep{HAWC_2021_pulsars}.
As most of the reported UHE sources to date are spatially coincident or in close proximity to powerful pulsars, disentangling the contribution from competing sources is a necessary step in current and future investigations.

Of the 12 LHAASO detected sources, \cite{LHAASO_PeV} reported the spectral energy density distributions for a selection of three: LHAASO\,J2226+6057, LHAASO\,J1825-1326 and LHAASO\,J1908+0621. For the latter source, a phenomenological fit comparing a hadronic and a leptonic model was presented. For the leptonic fit, the required power-law index of the electron injection spectrum was \num{-1.75}, with a cutoff energy of \SI{800}{\tera\electronvolt}. Such hard injection spectra are not generally expected, at least within a relativistic shock acceleration scenario. At first sight, this might further favour a hadronic model, however it was assumed that 
the magnetic energy density in the emission zone exceeded that of the photon field in their model. The other two sources are already listed in the HAWC 100~TeV source catalogue \citep{HAWC_UHE}, and consistent matches to the UHE emission using standard theoretical injection spectra were presented in \citetalias{Breuhaus}. It is thus important to explore if these results still apply in light of the improved data in the UHE regime, and also if such a scenario might also apply for new sources. It is of course possible, that the situation is more complex, with more than one source powering the detected emission \citep{HAWC_2021_J1825, Crestan_et_al_2021_J1908}. Naively, one might expect the likelihood that several nearby or overlapping sources accelerate particles to \si{\peta\electronvolt} energies is improbable, future high resolution instruments may be necessary to unambiguously determine if this is the case.

In this Letter, we revisit the predictions of \citetalias{Breuhaus} in light of the latest LHAASO results. We focus on the three sources for which SEDs are available, two of which overlap with HAWC sources previously considered in \citetalias{Breuhaus}.
In section \ref{sec:Method} we describe the methodology and section \ref{sec:Results} presents the results. At the end, we discuss our findings in section \ref{sec:Conclusion}.

\section{Method}\label{sec:Method}

To calculate the emission, we use a simple single-zone treatment.
In this model, accelerated electrons are injected over a time frame corresponding to the 
characteristic age of the associated pulsar 
into a region with constant homogeneous magnetic field and isotropic radiation fields. The injection rate is assumed to follow a time independent (isotropic) power-law distribution with exponential cut-off:
\begin{align}
    Q(E) = \dot{N}_0 \cdot \left(\frac{E}{\SI{1}{\erg}}\right)^{-\alpha} \cdot \exp\left(-\frac{E}{E_{\rm cut}}\right),
\end{align}
i.e. we will ignore the evolution of the source, and restrict our attention to the highest energy electrons, for which the cooling time is less than or close to the dynamical age.
We note that the impact of considering a finite source age rather than an equilibrium is modest in most cases, but improves the fit and modifies the best fit injection index in the case of younger sources and wide spectral coverage.

The total electron spectrum is calculated by taking into account energy losses due to both IC and Synchrotron emission. On timescales shorter than the dynamical time, adiabatic losses are assumed to be negligible. The solution can be obtained using a standard single-zone approach \citep[see e.g.][]{Atoyan_Aharonian_1999}, which has been implemented into the open source code GAMERA \citep{gamera}. 
Particle escape is neglected, a valid assumption if the particles diffuse sufficiently slowly, and cool fast. The energy is therefore deposited close to the source and the spectral data points obtained cover the entire emission region. 
We can check \emph{a posteriori} that these hold.

For the radiation fields, we use the large scale Galactic emission model of \cite{Popescu2017} together with the cosmic microwave background (CMB) at the respective source locations, assuming the sources are located at the position of the pulsar counterparts at their nominal distances. We retain the option to include a local enhancement factor of the diffuse photon field due to nearby photon sources / spiral arms (see \citetalias{Breuhaus}). Between source and detection at Earth, $\gamma$-rays are absorbed due to pair production with mostly far infrared (FIR) Galactic radiation fields and the CMB, the latter being important 
for energies above \SI{\sim 100}{\tera\electronvolt} \citep[e.g.][]{Gould78, Vernetto_Lipari_2016}. In the worst case, for the source LHAASO J1908+0621 associated to the pulsar PSR 1907+0632, the transmissivity at \SI{1}{\peta\electronvolt} is \SI{67}{\percent}.

\citet{LHAASO_PeV} provide a list of potential particle acceleration sites, including supernova remnants or pulsars, associated to the detected UHE sources. We focus here on the possible contribution from pulsars alone. 
For LHAASO\,J2226+6057 there is only one known high-power pulsar in the region of interest, while for the other sources two powerful pulsars are located within $\SI{1}{\degree}$ from the measured positions. 
The key physical parameters of these pulsars are provided in table \ref{tab:pulsars}, characteristic spin-down age is taken as a proxy for the true age of the pulsar.
In each case we then fit the normalisation, the injection index $\alpha$ and the cutoff energy $E_{\rm cut}$ (within physically motivated bounds) to the data for a fixed magnetic field for each pulsar association. The errors in determining $E_{\rm cut}$ are found to be highly asymmetric. We therefore estimate the \SI{95}{\percent} confidence interval separately using the probability density function of the $\chi^2$ distribution, and its dependence on $E_{\rm cut}$.

\begin{table}[]
    \centering
    \small
    \begin{tabular}{|c|c|c|c|c|}
    \hline
    \makecell{Pulsar\\(source)} & \makecell{Distance\\{[}\si{\kilo\pc}{]}}  & \makecell{Age\\{[}\si{\kilo\year}{]}} & \makecell{$L_{\rm s}$\\{[}\si{\erg\per\second}{]}} & \makecell{$u_{\rm ph}$\\{[}{\rm eV ~cm$^{-3}$}{]}} \\\hline\hline
  \makecell{PSR J2229+6114\\(J2226+6057)} &  \num{0.8}& \num{10.0}& \num{2.2e37} & 0.54 (0.16) \\\hline
\makecell{PSR J1826-1334\\(J1825-1326)} &  \num{3.1} &\num{21.4}& \num{2.8e36} & 2.15 (0.41)\\\hline
    \makecell{PSR J1826-1256\\(J1825-1326)} &  \num{1.6} & \num{14.4}& \num{3.6e36} & 1.21 (0.28)\\\hline 
    \makecell{PSR 1907+0602\\(J1908+0621)} & \num{2.4}  & \num{19.5} & \num{2.8e36} &  1.24 (0.28)\\\hline 
    \makecell{PSR 1907+0632\\(J1908+0621)} & \num{3.4}   & \num{11.3} & \num{5.3e35} &  1.50  (0.32)\\\hline
    \end{tabular}
    \caption{Adopted properties of the pulsars with possible association to the three LHAASO sources \citep[and references therein]{LHAASO_PeV}. The pulsar names, and associated LHAASO source (in brackets) are listed in the first column. $L_{\rm s}$ denotes the spin-down luminosity and $u_{\rm ph}$  is the integrated IR and optical/UV (in parentheses) photon energy density from the radiation model of \cite{Popescu2017} at the inferred pulsars' Galacto-centric radii and distances from the mid-plane. We distinguish the two contributions as photons with wavelength $\lambda \gtrless \SI{700}{\nano\meter}$, excluding CMB. }
    \label{tab:pulsars}
\end{table}

\section{Results}\label{sec:Results}

We apply the method described above to the sources LHAASO\,J2226+6057, LHAASO\,J1825-1326 and LHAASO\,J1908+0621, for which the spectra are provided in \citet{LHAASO_PeV}. 

\paragraph{LHAASO J2226+6057:} 

This source can be associated to the SNR G106.3+2.7 as well as the pulsar J2229+6114 and its nebula (the \lq\lq\emph{Boomerang}\rq\rq\, nebula) \cite[e.g.][]{Kothes}. The centroid of the $100$ TeV $\gamma$-ray source was also shown by the \citet{Tibet_J2226} to be spatially coincident with a molecular cloud, which might favour a hadronic origin. However, given the mature age of the SNR ($\lesssim 10^4$ years), rather unique conditions are necessary to motivate ongoing PeV acceleration at the SNR shock. We consider here the alternative scenario where the electrons/positrons are accelerated as part of the energy dissipation process of the pulsar wind, for example via shock acceleration \cite[e.g.][]{GiacintiKirk}.

\begin{figure}
    \centering
    \includegraphics[width=0.5\textwidth]{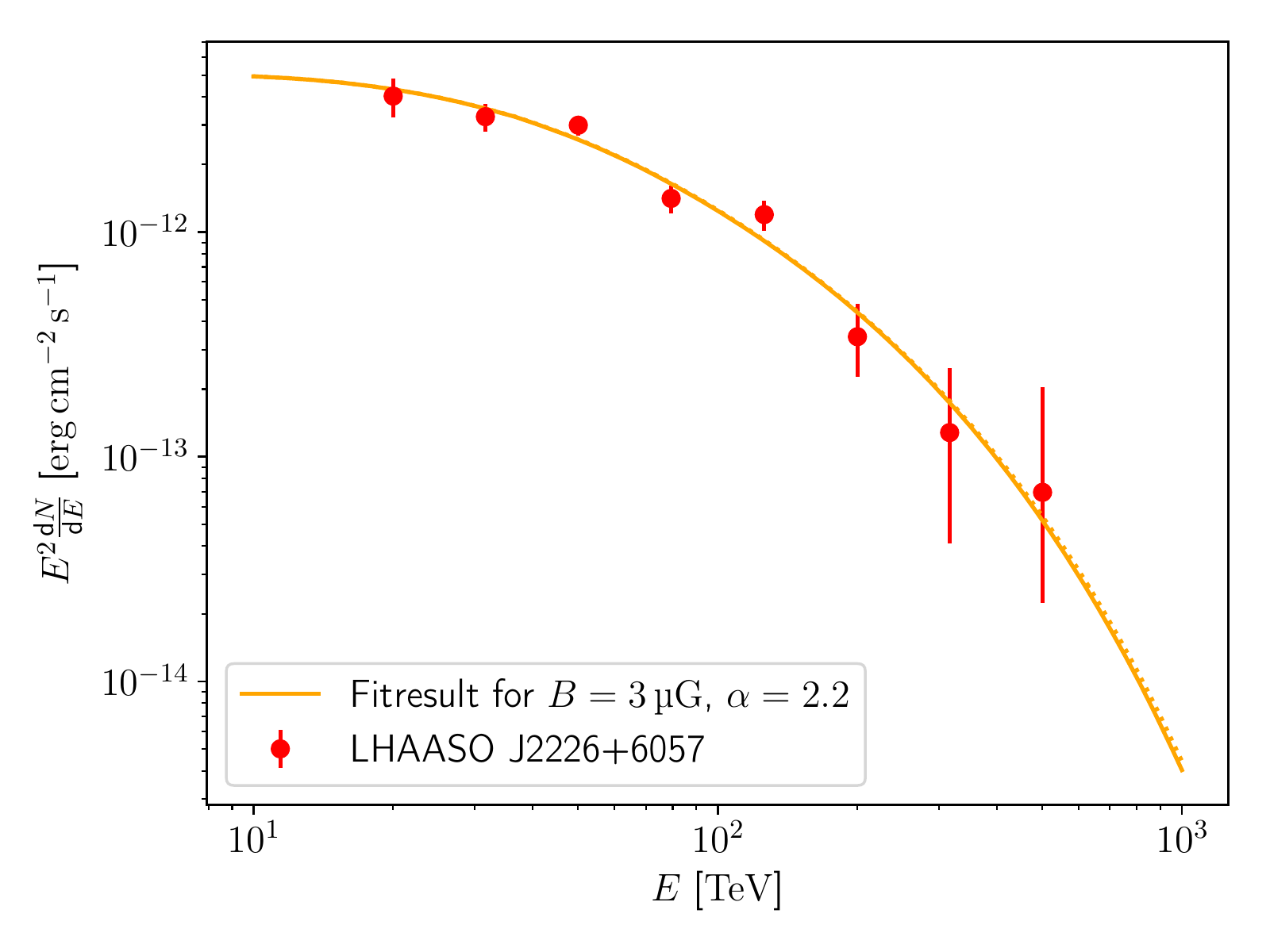}
    \caption{Data and models for LHAASO\,J2226+6057. The orange curve is a model fit for a magnetic field of $B = \SI{3}{\micro\gauss}$ and a fixed power-law index of the injection spectrum of $\alpha=2.2$, leaving $\dot{N}_0$ and $E_{\rm cut}$ as free parameters. The resulting values were $\dot{N}_0 = \SI{9e33}{\erg\per\second}$ and $E_{\rm cut} = \SI{420}{\tera\electronvolt}$. The dotted line shows the emission without absorption.}
    \label{fig:J2226}
\end{figure}

Using the pulsar parameters and photon field associated to the pulsar's position (see Table \ref{tab:pulsars}), fits based on the LHAASO data alone are not well constrained. While theoretical predictions for relativistic shock acceleration in general disfavour injection power-law indices $\alpha < 2$  \cite[e.g.][]{SironiReview}, equally good fits are possible over a broad range of $\alpha$, and cut-off energies. Figure \ref{fig:J2226} compares the LHAASO data with an exemplary model fit for $B=\SI{3}{\micro\gauss}$ and a fixed injection index $\alpha=2.2$. The resulting normalisation parameter is $\dot{N}_0 = \SI{9\pm2e33}{\per\erg\per\second}$ and the cutoff energy is $E_{\rm cut} = 420_{-130}^{+210}\,\si{\tera\electronvolt}$ at a \SI{95}{\percent} confidence level. The cut-off energy is comfortably below the maximum potential drop of the pulsar (see for example \citeauthor{AmatoOlmi21}~\citeyear{AmatoOlmi21} or \citetalias{Breuhaus} Appendix A).
The resulting flux corresponds to a fraction $\eta = \SI{0.13}{\percent}$ of the pulsar's spin-down power injected into electrons above \SI{1}{\tera\electronvolt}, or $\approx$ \SI{1}{\percent} for electrons above \SI{1}{\giga\electronvolt} assuming no break in the injected spectrum. PSR J2229+6114 is therefore easily able to provide the necessary power input.
Multi-wavelength models of the source have been explored by other authors, \citet{Liu20,Yu22}, where their best fits correspond to a $4~\mu$G magnetic field, with $\alpha=2.4-2.5$. Note that in both studies an IR photon field of $0.3$ eV cm$^{-3}$ was adopted, which differs from the model of \citet{Popescu2017} used here (see Table \ref{tab:pulsars}).

An injection spectrum of $\alpha=2.2$ is also consistent with the radio spectrum, $S_\nu \propto \nu^{-0.59}$, measured from the Boomerang nebula. \cite{Kothes} interpreted the break at $\nu \approx 5$ GHz, as an indication that this is in fact the cooled spectrum, though this would require a magnetic field strength of $2.6$~mG. Recent work has questioned whether such intense fields can exist on large scales \cite{Liu20}.  However, \cite{Kothes} themselves interpret the radio emission as originating in the \lq\lq crushed\rq\rq\, part of the nebula, due to reverse shock resulting from the SNR interaction with dense H\,\uppercase\expandafter{\romannumeral 1\relax} material in the north-east direction. In this scenario, the termination shock in the south west region would provide more favourable conditions not only for acceleration, but escape into the low density cavity evacuated by the SNR. This would also be consistent with the observed off-set between the emission centres of the Boomerang nebula and the LHAASO emission which is separated by $\approx$\SI{ 0.4}{\degree} to the south west. While a detailed numerical model is needed to explore the complex transport expected in such conditions, it is possible to make rough estimates of the expected diffusion coefficient to account for the redial extent of the LHAASO emission. We define a radial diffusion coefficient $D_{\rm r}= R^2/4 \tau$, where $R$ is the radius of the observed emission, and $\tau$ is the minimum of the system age and the total (energy dependent) cooling time. Adopting a half-angle of $\approx$ \SI{1}{\degree} \cite[see][Fig 1]{LHAASO_PeV} at a distance of $0.8$~kpc, we find for example at a representative energy of $100$TeV, $D_{\rm r}=\SI{1.5e27}{\square\centi\meter\per\second}$.
While this is smaller than the Galactic average of $ D_{\rm Gal} \approx 10^{26} E_{\rm GeV}^{1/3}~{\rm cm}^2{\rm s}^{-1} $, it is consistent with the Bohm limit for a $3~\mu$G field, $D_{\rm Bohm}(100~{\rm TeV}) = \SI{1.1e27}{\square\centi\meter\per\second}$.

LHAASO J2226+6057 is a source of considerable interest \cite[e.g.][]{Fujita,Geetal} with new multi-wavelength data forthcoming. Such broadband coverage will be essential to discriminate different interpretations regarding the origins of the UHE emission.

\begin{figure}
    \centering
    \includegraphics[width=0.5\textwidth]{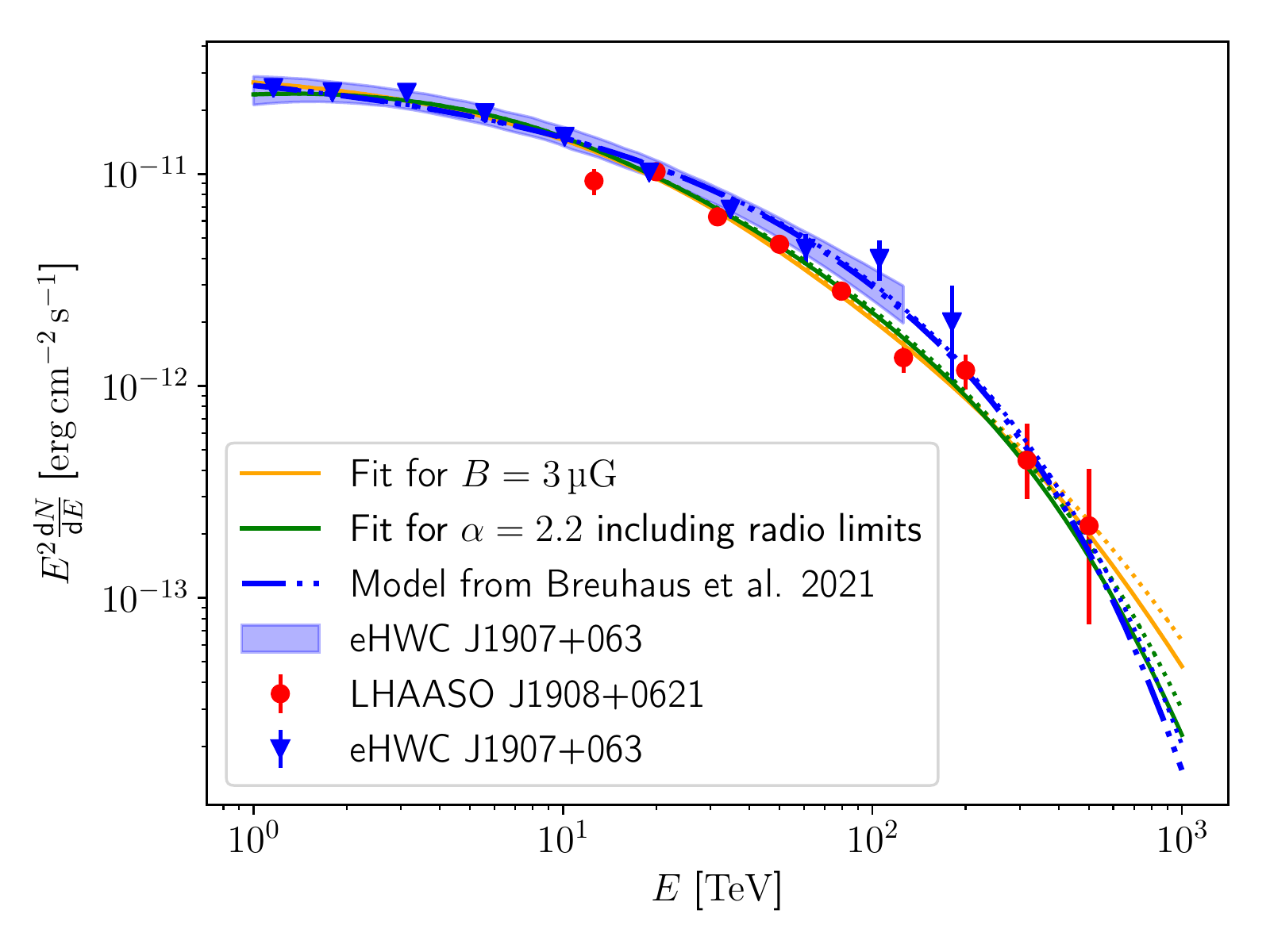}
    \caption{Model fit for LHAASO\,J1908+0621 and eHWC\,J1907+063 together with the corresponding data and the model from \citetalias{Breuhaus} for eHWC\,J1907+063. Red data points show the data from the LHAASO collaboration \citep{LHAASO_PeV} and the blue triangles data from the HAWC collaboration for eHWC\,J1907+063 \citep{HAWC_UHE}. The blue band shows the systematic and statistical error for eHWC\,J1907+063 \citep{HAWC_UHE} and the blue dashed-dotted line the corresponding model from \citetalias{Breuhaus} extended to higher energies and corrected for absorption. The orange curve represents the joint fit to the data from HAWC and LHAASO for $B = \SI{3}{\micro\gauss}$, resulting in $\dot{N}_0 = \SI{1e36}{\per\erg\per\second}$ and $\alpha = 2.66$, assuming an association to PSR\,1907+0602. $E_{\rm cut}$ was only constrained by the fit to be larger than \SI{1.4}{\peta\electronvolt} and was set to equal the Hillas limit of \SI{4.2}{\peta\electronvolt}. The green solid line shows the result from the fit including the radio upper limits with the $B$-field as a free parameter but a fixed value for $\alpha = 2.2$. The dotted lines in the corresponding colors show the emission without absorption.}
    \label{fig:J1908}
\end{figure}

\paragraph{LHAASO J1908+0621:} 
This source overlaps with the well known VHE source \cite[e.g.][]{HESS_1908} and UHE source eHWC\,J1907+063 \citep{HAWC_UHE}. However, since two powerful pulsars are spatially coincident with the LHAASO source, we must consider them separately. 
We perform a joint fit using the new data from the LHAASO observatory together with the HAWC data. The result for $B = \SI{3}{\micro\gauss}$ associated to the pulsar PSR\,1907+0602 is shown in figure \ref{fig:J1908} with the orange line. The red data points show the data from LHAASO \citep{LHAASO_PeV}, and the blue triangles the data from HAWC \citep{HAWC_UHE}. We also show the systematic and statistical error band from HAWC as a blue shaded region. 

The resulting fit parameters for a PSR\,1907+0602 association are: $\dot{N}_0 = \SI{10\pm 2e35}{\per\erg\per\second}$, $\alpha = 2.66 \pm 0.03$.
 
The best fit cutoff is $E_{\rm cut} = \SI{10}{\peta\electronvolt}$, but any value above \SI{1.4}{\peta\electronvolt} is consistent with the data at \SI{95}{\percent} confidence level. The Hillas limit for a pulsar with the spoin-down power of  PSR 1907+0602 is \SI{4.2}{\peta\electronvolt}. In figure \ref{fig:J1908}, this value was chosen for $E_{\rm cut}$.
This scenario requires $\sim$40\% of the spin-down power of the pulsar in electrons above 1 TeV. To avoid exceeding the available energy budget, electrons should be injected at a minimum energy of \SI{240}{\giga\electronvolt}. However, if a break in the injection spectrum  below \SI{1}{\tera\electronvolt} exists, this condition can be relaxed. Data in the \si{\giga\electronvolt} to \si{\tera\electronvolt} regime can help to detect possible spectral breaks, and evidence for such a break exists \citep{Crestan_et_al_2021_J1908}.
For an association with PSR\,1907+0632, the values $\dot{N}_0 = \SI{4.2\pm 0.5e36}{\per\erg\per\second}$ and $\alpha = 2.79\pm 0.03$ are obtained. The cutoff energy is constrained to be $>\SI{2}{\peta\electronvolt}$ at \SI{95}{\percent} confidence level. In addition to being in tension with the Hillas limit of \SI{1.8}{\peta\electronvolt}, to account for the emission approximately 7 times available pulsar power must be transferred to electrons above \SI{1}{\tera\electronvolt}. It is therefore highly unlikely that PSR\,1907+0632 alone can account for this source, even allowing for evolution of the injected power over the cooling timescale.

We also compare the fit here to the equilibrium model from \citetalias{Breuhaus} shown as the blue dashed dotted line in figure \ref{fig:J1908}. The curve, developed initially to fit just the HAWC data, provides a reasonable match to the ultra high energy emission detected by LHAASO. Both models are nearly identical at low energies, but diverge slightly at higher energies due to the higher cutoff energy in the new model. A more gradual turnover at the highest energies is needed to match the slightly reduced flux values of the LHAASO relative to HAWC data in the overlapping energy range. As shown previously \citepalias{Breuhaus}, data from the IRAS survey excludes enhancement in the FIR radiation fields in the range from \SIrange{60}{100}{\micro\meter} and the radiation fields cannot be very different from the average Galactic field, leaving the magnetic field as the only uncertain environmental parameter. The fit parameters for larger magnetic field values result in smaller $\alpha$ and smaller $E_{\rm cut}$. Insisting that $\alpha \gtrsim 2.0$ implies an upper limit on the magnetic field strength of \SI{\sim 8}{\micro\gauss}, with a resulting cutoff energy of $730_{-310}^{+810}\,\si{\tera\electronvolt}$ at \SI{95}{\percent} confidence level. However, additional physical effects such as energy dependent confinement could in principle allow injection spectra harder than $\alpha = 2.0$ to occur, and should not be ruled out \emph{a priori}.

For $B = \SI{3}{\micro\gauss}$ the cutoff energy is constrained to be very close to 1~PeV. In scenarios with larger $B$-fields, the value for $E_{\rm cut}$ is reduced, but also less well constrained. Even for $B = \SI{8}{\micro\gauss}$, a cutoff energy of \SI{1}{\peta\electronvolt} is still consistent within errors.

A detection of radio emission from the source region can help to constrain the magnetic field, but no emission on the scales of the LHAASO source extend was detected so far. \cite{Duvidovich_2020_J1908} calculated upper limits at \SI{1.5}{\giga\hertz} and \SI{6}{\giga\hertz} based on the non-detection of \SI{1}{\milli\jansky\per\beam} (\SI{10}{\micro\jansky\per\beam}) for a beamsize of $\SI{51.1}{\arcsecond}\times\SI{39.5}{\arcsecond}$ ($\SI{12.2}{\arcsecond}\times\SI{8.6}{\arcsecond}$). With an estimated radial size of \SI{1}{\degree} \citep[][figure 1]{LHAASO_PeV} these limits translate into upper limits of \SI{3.0e-13}{\erg\per\square\centi\meter\per\second} (\SI{2.3e-13}{\erg\per\square\centi\meter\per\second}). The model is not necessarily consistent with these limits, but if electrons are indeed injected only above \SI{240}{\giga\electronvolt} to fulfil the energy requirement, or a break in the injection spectrum below \si{\tera\electronvolt} energies exists, the emission is below the upper limit. Models with lower minimum injection energy and without a break fulfilling the upper limit exist as well: In a different approach, we required consistency with the radio limits and performed a fit with the $B$-field as a free parameter, but a fixed value for $\alpha = 2.2$. This leads to $B = \SI{5.7\pm 0.3}{\micro\gauss}$, $\dot{N}_0 = \SI{2.8\pm 0.2 e35}{\per\erg\per\second}$ and $E_{\rm cut} = 830_{-220}^{+360}\,\si{\tera\electronvolt}$. In this scenario, \SI{30}{\percent} of the spin-down power is injected into electrons above \SI{1}{\tera\electronvolt} and a minimum injection energy of \SI{10}{\giga\electronvolt} is required to be consistent with energy constraints.

The diffusion coefficient for $B = \SI{3}{\micro\gauss}$ at \SI{100}{\tera\electronvolt} for a source size of \SI{1}{\degree} is \SI{1.3e28}{\square\centi\meter\per\second}, which is an order of magnitude above the value for Bohm diffusion (\SI{1.1e27}{\square\centi\meter\per\second}). For $B = \SI{5.8}{\micro\gauss}$ used in the second model, the diffusion coefficient is \SI{3.9e28}{\square\centi\meter\per\second}, again far above the corresponding Bohm diffusion coefficient of \SI{5.8e26}{\square\centi\meter\per\second}.\\

\begin{figure}
    \centering
    \includegraphics[width=0.5\textwidth]{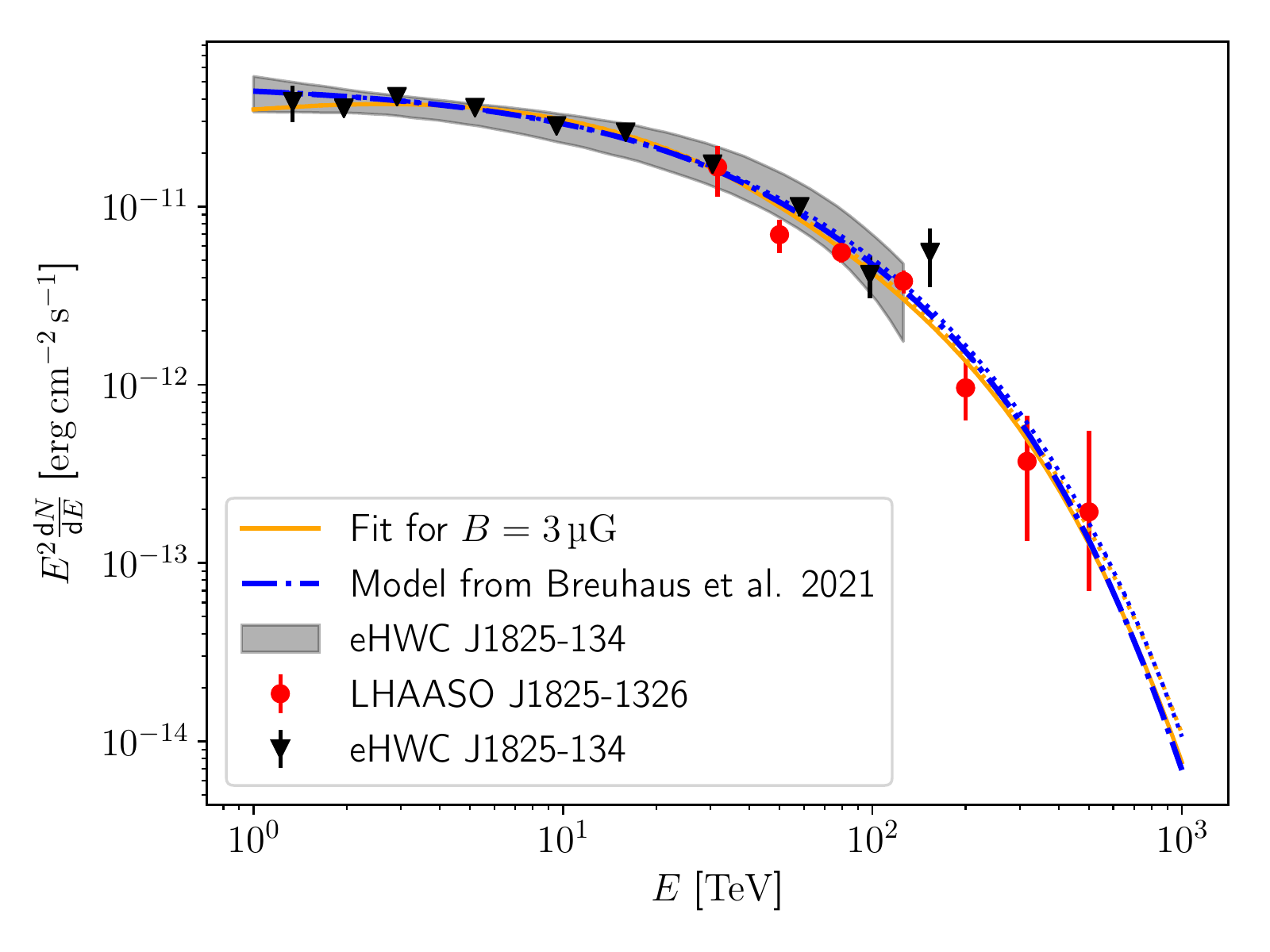}
    \caption{Model fit for LHAASO\,J1825-1326 and eHWC\,J1825-134 together with the corresponding data and the model from \citetalias{Breuhaus} for eHWC\,J1825-134. Red data points show the data from the LHAASO collaboration \citep{LHAASO_PeV} and the black triangles data from the HAWC collaboration for eHWC\,J1825-134. The blue band shows the systematic and statistical error for eHWC\,J1825-134 \citep{HAWC_UHE} and the green dashed line the corresponding model from \citetalias{Breuhaus} extended to higher energies and corrected for absorption. The orange curve represents the joint fit to the data from HAWC and LHAASO for $B = \SI{3}{\micro\gauss}$, resulting in $\dot{N}_0 = \SI{5e35}{\per\erg\per\second}$, $\alpha = 2.23$ and $E_{\rm cut} = \SI{390}{\tera\electronvolt}$. The fit result for the association to PSR\,J1826-1256 is nearly identical to the orange curve. The dotted lines in the corresponding colors show the emission without absorption.}
    \label{fig:J1825}
\end{figure}

\paragraph{LHAASO J1825-1326:} The environment of this source is equally complex. As with  LHAASO\,J1908+0621, it also possesses a HAWC counterpart, and a similar joint fit was performed. The magnetic field in the electron cooling zone was again initially fixed to be \SI{3}{\micro\gauss}. Two candidate pulsars are potentially associated to LHAASO\,J1825-1326: PSR\,J1826-1334 and PSR\,J1826-1256. For an association to PSR\,J1826-1334 the fit gives $\dot{N}_0 = \SI{5\pm 2e35}{\per\erg\per\second}$, $\alpha = 2.23\pm 0.09$ and $E_{\rm cut}$ is constrained to  $390_{-150}^{+340}\,\si{\tera\electronvolt}$ at a \SI{95}{\percent} confidence level. This model corresponds to $\eta = \SI{50}{\percent}$ and is shown in Figure~\ref{fig:J1825} with the orange curve. A minimum injection energy of \SI{90}{\giga\electronvolt} is required to not exceed the available power input from the pulsar, but a spectral break below \si{\tera\electronvolt} energies relaxes this condition. The cutoff energy of the model is consistent with the Hillas limit of \SI{4.2}{\peta\electronvolt}. The $\gamma$-ray emission from the fit for an association to PSR\,J1826-1256 is nearly identical for the energy range shown, the resulting fit parameters in this case are $\dot{N}_0 = \SI{4\pm 2e35}{\per\erg\per\second}$, $\alpha = 2.39\pm 0.09$ and $E_{\rm cut} =410_{-160}^{+380}\,\si{\tera\electronvolt}$ at a \SI{95}{\percent} confidence level, with corresponding efficiency $\eta = \SI{20}{\percent}$. A minimum injection energy of \SI{25}{\giga\electronvolt} is required if no spectral break below \SI{1}{\tera\electronvolt} exists. The cutoff energy is below the Hillas limit of \SI{4.7}{\peta\electronvolt}.

The model from \citetalias{Breuhaus} for the HAWC source eHWC\,J1825-134 is depicted in figure \ref{fig:J1825} with a dashed dotted blue curve and follows closely the model from the joint fit. While the $B$-field was the same in both cases, the radiation fields were not: In \citetalias{Breuhaus} enhancement factors of 3 and 5 were used for the association to PSR\,J1826-1334 and PSR\,J1826-1256, respectively. Due to the presence of a cooling break in the new model, no enhancement is needed. Fits for higher $B$-field value decrease the resulting value of $\alpha$. For example, For $B = \SI{5}{\micro\gauss}$ and an association to PSR J1826-1334 the resulting parameters are $\dot{N}_0 = \SI{1.8\pm 0.8e35}{\per\erg\per\second}$, $\alpha = 1.9\pm 0.1$ and $E_{\rm cut} = 310_{-110}^{+200}\,\si{\tera\electronvolt}$.
An enhancement in the radiation field will increase the resulting value for $\alpha$. As shown in \citetalias{Breuhaus}, enhancements of the IR and UV fields up to a factor of 16 are compatible with upper-limits from IRAS data, and therefore a variety of models for larger $B$-field values are viable as long as the source-specific radiation fields are not constrained further.

As for LHAASO J1908+0621, no radio emission on the scales of the LHAASO source extend was detected so far. \cite{Duvidovich_2019_J1825} calculated an upper limit at \SI{1.4}{\giga\hertz} based on the non-detection of \SI{0.24}{\milli\jansky\per\beam} for a beamsize of $\SI{9.24}{\arcsecond}\times\SI{6.43}{\arcsecond}$. With an estimated radial size of \SI{1}{\degree} given by \cite{LHAASO_PeV} this limit translates into an upper limit of \SI{2.3e-12}{\erg\per\square\centi\meter\per\second}. The models developed above fulfil this constraint.

The diffusion coefficients at \SI{100}{\tera\electronvolt} are \SI{2.3e28}{\square\centi\meter\per\second} (PSR J1826-1334) and \SI{6.0e27}{\square\centi\meter\per\second} (PSR J1826-1256), both above the Bohm diffusion case of \SI{1.1e27}{\square\centi\meter\per\second}.\\

As shown above, reasonable leptonic single-zone models exist for all of the sources. Due to the uncertainty in the environmental conditions, considerable flexibility in the fitting models is permitted. In general however, larger magnetic field values require harder injection spectra, while enhanced radiation fields have the opposite effect. To constrain parameters and to confirm or rule out different scenarios, it is therefore necessary to have either more information about the specific environmental conditions, or to include multi-wavelength data in the modelling~\footnote{Noting that great care is needed in combination of data on different angular scales and/or variable surface brightness sensitivity, to avoid reaching inappropriate conclusions}. This requires a detailed source-specific investigation. For each source, multiple possible counterparts exist and taking into account different sources might be necessary, especially at energies below several \si{\tera\electronvolt}. In all models, the cutoff energies were constrained to be between few 100 \si{\tera\electronvolt} to more than a \si{\peta\electronvolt} in some cases. Nearby pulsars can supply sufficient power to account for the observed flux levels, and therefore the nebulae of high spin-down power pulsars may frequently house particles with energies in excess of \SI{100}{\tera\electronvolt}. The estimated diffusion coefficients are orders of magnitudes below the interstellar CR diffusion coefficients, but similar to estimates for the region
around Geminga \citep{HAWC_Geminga}. Such low diffusion coefficients could be caused by self-confinement \citep{Evoli_et_al_2018}.

\section{Conclusion}\label{sec:Conclusion}
In this Letter we have shown, that the emission from LHAASO\,J2226+6057, LHAASO\,J1825-1326 and LHAASO\,J1908+0621 is consistent with simple one zone models of inverse Compton emission, provided that, in the case of LHAASO\,J2226+6057 and LHAASO\,J1825-1326, the magnetic fields in these systems are somewhat lower than typical Galactic average values, or local radiation field energy densities somewhat higher than the large-scale average.
For each source there exists at least one plausibly associated pulsar providing the necessary power to account for the observations above \SI{1}{\tera\electronvolt}. 
This fact, coupled with the validity of the IC emission scenario as demonstrated here, can be seen as strong evidence that high spin-down power pulsars routinely accelerate electrons to hundreds of \si{\tera\electronvolt} and even \si{\peta\electronvolt} energies. The detection of PeV $\gamma$-rays from the Crab nebula, recently reported by LHAASO, supports this claim \citep{LHAASO_Crab}. Since all such neutron stars are thought to launch an ultra-relativistic wind, such extreme accelerators may be surprisingly common, provided they satisfy the associated Hillas condition \citep{Hillas1984}. However, as discussed in \citetalias{Breuhaus}, special conditions may be necessary for these sources to produce detectable fluxes.

Regions with strong radiation fields and/or low magnetic-fields exist in many parts of the Galaxy \cite[e.g.][]{ICRC2021}, providing suitable conditions for hard inverse Compton emission at energies above \SI{100}{\tera\electronvolt}.
Radiation dominated sources may be particularly prevalent close to the Galactic centre region \cite[e.g.][]{HintonAharonian} where the energy density in the diffuse photon field increases rapidly with decreasing Galacto-centric radius \citep{Popescu2017}.
Future detection of new UHE $\gamma$-ray sources are anticipated, with many new results expected from the LHAASO observatory.
Future southern hemisphere observatories such as CTA South and SWGO, which are better positioned to observe the Galactic plane and central region, are highly desirable in this effort. 

The search for a firm identification of the sources of PeV protons and nuclei is ongoing.
To confirm or rule out different scenarios, it is crucial to have detailed information about the environmental conditions and to take multiwavelength data into account. In some cases, it may be necessary to consider multi-source models. Very high energy $\gamma$-ray spectral information alone is in most cases not sufficient to discriminate between leptonic and hadronic scenarios, but we note that \emph{spatially resolved} VHE-UHE emission might well be, in particular if the energy-dependent morphology implied by the cooling limited IC picture presented here can be demonstrated. The CTA project is particularly important in this regard~\citep{CTA_sciencecase_2019}.

\section{Acknowledgements}
The authors thank F. Aharonian for valuable comments.
For the numerical calculations, we made use of the open source GAMERA code \citep{gamera}. We also used the python packages SciPy \citep{SciPy}, NumPy \citep{NumPy}, Astropy \citep{astropy:2013, astropy:2018} and Matplotlib \citep{Matplotlib}.

\bibliographystyle{aa}
\bibliography{references}

\end{document}